# CMOS-compatible, piezo-optomechanically tunable photonics for visible wavelengths and cryogenic temperatures

**P. R. Stanfield, A. J. Leenheer, C. P. Michael, R. Sims, and M. Eichenfield***

*Sandia National Laboratories, P.O. Box 5800 Albuquerque, New Mexico 87185, USA*

*\*meichen@sandia.gov*

**Abstract:** We demonstrate a platform for phase and amplitude modulation in silicon nitride photonic integrated circuits via piezo-optomechanical coupling using tightly mechanically coupled aluminum nitride actuators. The platform, fabricated in a CMOS foundry, enables scalable active photonic integrated circuits for visible wavelengths, and the piezoelectric actuation functions without performance degradation down to cryogenic temperatures. As an example of the potential of the platform, we demonstrate a compact (~40 μm diameter) silicon nitride ring resonator modulator operating at 780 nm with intrinsic quality factors in excess of 1.5 million, >10 dB change in extinction ratio with 2 V applied, a switching time less than 4 ns, and a switching energy of 0.5 pJ/bit. We characterize the exemplary device at room temperature and 7 K. At 7 K, the device obtains a resistance of approximately $2x10^{13}$ Ohms, allowing it to operate with sub-picowatt electrical power dissipation. We further demonstrate a Mach-Zehnder modulator constructed in the same platform with piezoelectrically tunable phase shifting arms, with 750 ns switching time constant and 20 nW steady-state power dissipation at room temperature.

## 1. Introduction

Many application spaces with operational wavelengths well below the silicon bandgap of ~1100 nm and requiring operation at cryogenic temperatures have an unmet need for a scalable, low-power, active photonic integrated circuit architecture that would allow for dynamical optical routing and optical phase and amplitude modulation. The need for scalability would be best satisfied using a completely CMOS compatible set of materials and fabrication methods. In addition, this would facilitate integrated driver electronics, which are requisite for the electrical I/O density of many applications when implemented at scale. Yet, to date, such a platform has not been effectively demonstrated.

A modest number of examples of CMOS compatible modulators that could satisfy all these requirements exist in the literature. For example, there are many demonstrations of electrostatically actuated optical structures in CMOS compatible material sets [1-5]; however, *most* of these are done in silicon photonic structures, and silicon is obviously incompatible with operation at visible wavelengths. A more modest number of electrostatically actuated silicon nitride photonic devices do exist [6, 7]. Regardless, electrostatic actuation is difficult to use for anything but small-signal modulation, as it is intrinsically nonlinear and prone to collapse (pull-in) on account of the nature of the electrostatic force of a deformable capacitor. In addition, they are generally unsuitable for high-speed modulation; this arises due to the necessity of using large capacitances to generate large electrostatic forces, resulting in large RC time-constants. The most compelling CMOS compatible modulator devices presented have used direct electro-optic modulation of aluminum nitride (AlN) waveguiding layers [8-10]. Aluminum nitride can be used in CMOS fabrication with piezoelectric and electro-optic properties approximating single crystal films when grown by physical vapor deposition (PVD) [11]; however, these AlN films suffer from high optical losses at short wavelengths due to Rayleigh scattering arising from the polycrystallinity of aluminum nitride grown by PVD [10, 12, 13]. When applied to electro-optic modulation applications, this wavelength-dependent loss, combined with the large $LV_\pi$ characteristic of aluminum nitride, results in fairly poor performance at near-visible wavelengths and shorter. Another solution in the literature is thermo-optic tuning of CMOS compatible dielectrics such as silicon nitride and silicon dioxide. Thermo-optic modulators suffer from a relatively high power consumption necessary to produce a sufficient refractive index change that limits their potential scalability. This problem is compounded when attempting to operate at cryogenic temperatures where the efficacy of temperature tuning decreases and thermal budgets become even more constrained [14].

There have been attempts to use aluminum nitride to piezoelectrically strain optical waveguides to achieve modulation. This form of modulation allows a larger range of waveguide materials, with previous demonstrations of AlN piezoelectrically modulating AlN waveguides directly [15-21], and silicon nitride waveguides [22-24]. Demonstrations of directly piezoelectrically modulating AlN waveguides suffer from the same high losses at near-visible and shorter wavelengths as mentioned above. The existing demonstrations of aluminum nitride-based piezoelectric modulation of silicon nitride [22-24] have only been used for telecom c-band modulation, used high-temperature silicon nitride films that are not CMOS back-end-of-line compatible, and suffered from a combination of high optical loss and poor piezoelectric responsivity (for example, more than 10x smaller frequency shift per volt for a ring resonator, when compared to this work); this ultimately results in significantly poorer piezo-optomechanical performance as compared to the demonstration herein.

Potentially wavelength agnostic platforms such as PZT strain modulation have been presented previously [25-30], but are not strictly CMOS compatible. Beyond this fabrication constraint, PZT is challenging to use because of significant hysteresis and an extremely large dielectric constant, which limits modulation speed. More exotic modulation platforms such as $LiNbO_3$, polymer waveguides, and others have demonstrated impressive results [31-35], but, lacking CMOS compatibility, they do not have a clear path towards high-yield, very large-scale integration, which is critical for many applications.

In this work, we present a novel active silicon nitride photonic integrated circuit platform, fabricated in a CMOS foundry, that is piezo-optomechanically actuated by a tightly mechanically coupled aluminum nitride actuator layer. This allows the system to operate, at least in principle, over the transparency window of silicon nitride, which is approximately 300 nm to 6 μm [36, 37]. As an example of the potential of the platform, we demonstrate a compact (~40 μm diameter) silicon nitride ring resonator modulator operating at 780 nm with intrinsic quality factors in excess of 1.5 million, >10 dB change in extinction ratio with 2 V applied, a switching time less than 4 ns, and a switching energy of 0.5 pJ/bit. We characterize the exemplary device at room temperature and 7 K, with no degradation in the optical or piezoelectric properties over that range. At 7 K, the device obtains a resistance of approximately $2 \times 10^{13}$ Ohms, allowing it to operate with 0.2 picowatt electrical power dissipation when holding the device at its minimum insertion loss with 2 V, making it highly suitable for dense active photonic integrated circuits operating in cryogenic systems where dissipated electrical power must be removed with limited cooling power. We further demonstrate a Mach-Zehnder modulator constructed in the same platform with piezoelectrically tunable phase shifting arms, with 750 ns switching time constant and 20 nW steady-state power dissipation at room temperature.

## 2. Device Architecture

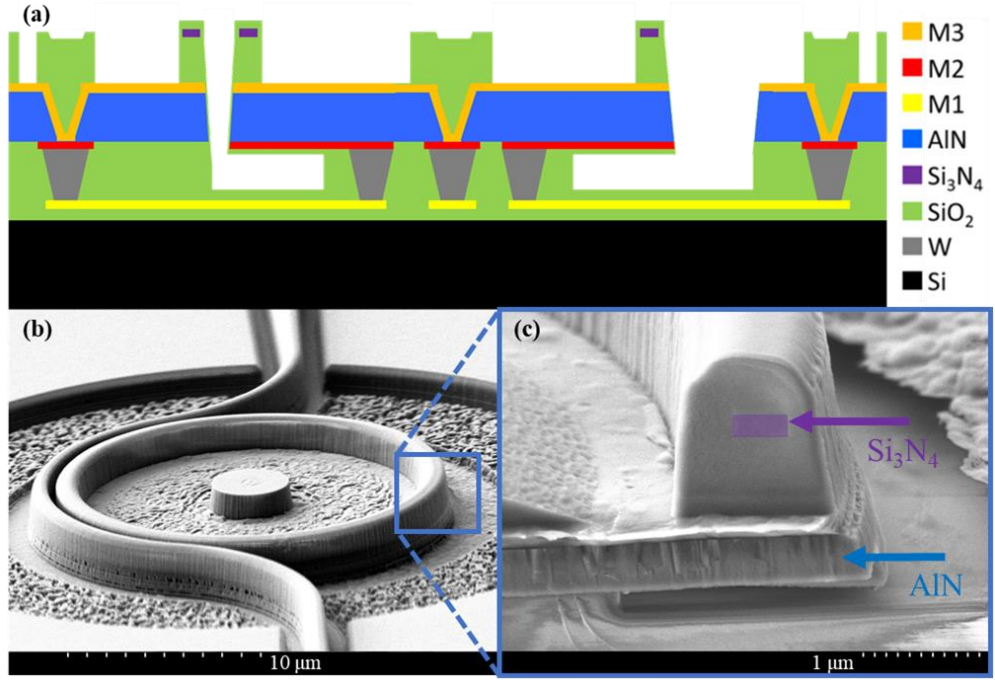

Fig. 1. Device architecture: (a) Schematic cross-section of a modulator device. (b) Scanning electron micrograph of fabricated ring modulator device. (c) Cross-sectional scanning electron micrograph of fabricated ring modulator device showing false-colored silicon nitride waveguide in purple transparent box.

As shown in Fig. 1(a), the device architecture uses silicon nitride photonic elements clad in silicon dioxide, tightly mechanically coupled to an aluminum nitride actuator with patterned metal electrodes above and below that allow for a vertical field to be applied to the piezoelectric AlN film; this allows for the use of the piezoelectric coupling coefficients, $d_{33} = 5$ pm/V and $d_{31} = -2$ pm/V. Aluminum nitride has been utilized in many low-power piezoelectric MEMS [38]. It has exceptionally low leakage current and can be deposited via CMOS-compatible, low temperature sputtering. A silicon dioxide layer above the AlN's top electrode provides an optical buffer for the ultra-low-loss silicon nitride photonics layer [39] as seen in Fig. 1(c). An aluminum routing metal layer sits under the aluminum nitride's bottom electrode, separated by a thick silicon dioxide spacer. Tungsten vias connect all three metal layers and allow the fields to be applied via a low-loss electrical connection that can be made with surface electrodes at any location on the chip. The tungsten electrical vias on the surface are protected by a silicon dioxide cap, as shown in Fig. 1(b) in the center of the device. Thus, the modulator platform is built on what is normally used as the top metal layer of a CMOS back-end-of-line process, allowing straight-forward post-CMOS integration for driver circuits, which could ultimately allow incredibly dense electrical I/O to occur on-chip, with only modest electrical connections off-chip [40]. A patterned amorphous silicon release layer is buried in silicon dioxide, underneath the AlN and its lower metal electrode, allowing a precise and terminal undercut later in the process and enabling accurate targeting of specific mechanical resonance characteristics.

Modulation is achieved via piezo-optomechanical coupling: i.e., optomechanical and photoelastic coupling in the silicon nitride photonics, driven piezoelectrically by applying electric fields to the AlN. The piezoelectrically generated strain induces photoelastic refractive index changes in the silicon nitride and silicon dioxide cladding, as well as movement of material boundaries that causes optomechanical changes to the effective refractive index of the optical mode [41]. These two effects combine to form an effective voltage-induced refractive index shift. In the case of the ring modulator, the effective *responsivity* of the device is characterized in terms of df/dV where f is the optical mode's resonant frequency. For the MZI modulators, responsivity is characterized by dΦ/dV where Φ is the differential phase shift between the two arms of the modulator when opposite drive voltages are applied.

Both the bandwidth and the responsivity of the modulator are largely determined by the mechanical design of the aluminum nitride actuator, specifically the mechanical compliance. The mechanical response is quasi-static below the lowest mechanical eigenfrequency, placing an upper bound on the modulator bandwidth. There is an inverse relationship between the deformation, and thus the responsivity, of the modulator and its bandwidth. This leads to a design trade-off, with stiffer designs having higher mechanical eigenfrequencies but actuating a correspondingly smaller amount.

Previously demonstrated strain modulators have relied on actuating a waveguide buried in a thick oxide buffer layer. In our architecture, the waveguide is tightly coupled to the actuation platform with the minimum possible oxide to sufficiently buffer the optical mode form the metal below. As seen in Fig. 1(c), excess oxide is etched away from the top of the platform to increase compliance and the ring is positioned close to the edge of the actuation platform to maximize waveguide strain. The responsivity is also significantly higher than similar devices previously demonstrated in aluminum nitride [22-24], which used an 800 nm thick and 1.8µm wide silicon nitride waveguide with 2 µm oxide separating the ring from the actuator.

*2.1 Ring Modulator Design*

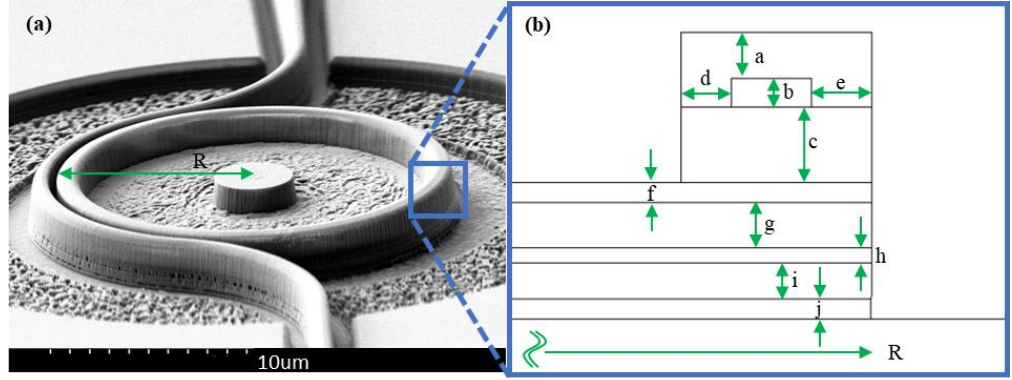

Fig. 2. (a) Representative device for ring modulator design. (b) Nominal device architecture cross section with labels for critical layer dimensions.

The ring modulator design, shown in Figs. 2(a) and 2(b), is based on the mechanical deformation of a silicon nitride ring with silicon dioxide cladding that is evanescently coupled to a waveguide formed in the same waveguide layers, ideally with close to critical coupling for large modulation depths. The ring sits on an aluminum nitride disk actuator of approximately the same diameter, buffered from the metal electrodes of the disk actuator by a thick silicon dioxide bottom cladding.

Table 1. Nominal Ring Modulator Device Parameters

| Label | a | b | c | d | e | f | g | h | i | j | R |
|---|---|---|---|---|---|---|---|---|---|---|---|
| Material | SiO$_2$ | Si$_3$N$_4$ | SiO$_2$ | SiO$_2$ | SiO$_2$ | Al | AlN | Al | SiO$_2$ | Si | N/A |
| Length (µm) | 0.460 | 0.285 | 0.750 | 0.500 | 0.600 | 0.200 | 0.450 | 0.150 | 0.360 | 0.200 | 20.4 |

The resonant frequency of the $m$ th optical mode of the ring is given by $f = mc/(2\pi R_{eff} n_{eff})$, where $m$ is the azimuthal eigenvalue of the mode, $R_{eff}$ is the effective radius of that mode, and $n_{eff}$ is the effective refractive index of the mode. Figures 3(a) and

3(b) shows the cross-sectional deformation in a finite element simulation of a ring modulator device when a voltage is applied across the AlN.

The performance of the devices is determined to a large extent by the thicknesses of all the layers in the system, as well as the in-plane dimensions of the features defined in these layers. The layer thicknesses were determined by optimizing performance via finite element multiphysics modeling in combination with myriad fabrication constraints. Of those constraints, the most important two follow. First, the etch which defines the ring modulator and mechanically separates the actuation platform from the surrounding material is difficult due to its high aspect ratio: approximately 2 µm etch depth and 200 nm etch width. The aspect ratio limits plasma etchant energies at the bottom of the etched trench and places an upper bound on the total thickness of the modulator stack. Second, the piezoelectric coefficients of polycrystalline aluminum nitride films are strongly dependent on thickness, with a rapid degradation of the coefficients (and thus the device responsivity) below a film thickness of approximately 300 nm [11]. While some of this performance modeling and optimization will be described below, for the purposes of this work we will focus on describing variations around a baseline design specified in Fig. 2(b) and Table 1.

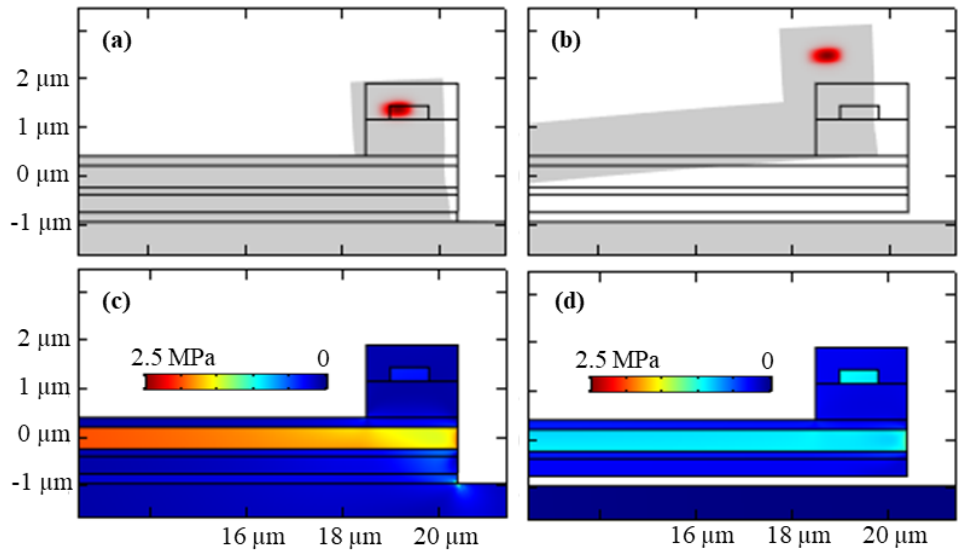

Fig. 3. (a) An axisymmetric finite element analysis of the TE00 optical resonance being piezo-optomechanically actuated in a non-undercut ring modulator at -1 V. Deformation is scaled by 10000x. The optical power circulating the ring resonator is shown in red. (b) An image of the actuation in a released ring modulator with the same applied voltage and scaling. (c) Von Mises stress from an applied voltage of -1V to a non-undercut ring modulator. (d) Von Mises stress from an applied voltage of -1V to an undercut ring modulator.

When the ring is deformed by actuation of the underlying aluminum nitride disk, both the radius, $R_{eff}$, and the effective index, $n_{eff}$, of the resonator mode change, shifting the resonant frequency. Figures 3(a) and 3(b) shows the scaled geometric deformation which modifies the path length of the ring modulator, while Figs. 3(c) and (d) shows stress from the actuation of the ring which modifies the effective index. The ring deforms in both the radial and vertical directions, as shown in Fig. 4(a), but the optomechanical/geometric contribution to the responsivity depends only on the radial component, to first order in the deformation. The ratio of radial to vertical deformation of the ring differs depending on the exact details of the modulator, including the size of the actuator and ring, oxide cladding thicknesses, all other film thicknesses, film compositions (which can vary the mechanical properties of the films) and fabrication details including residual stresses and etch angles. A non-undercut ring modulator device remains relatively compliant in the radial direction while being securely anchored to the substrate, largely eliminating vertical deformation. These mechanical properties give the device a relatively large responsivity for its stiffness compared to undercut devices, as seen in Fig 4(b). When the same device is undercut, the stiffness significantly decreases, improving the actuation, but a significant portion of the induced deformation is vertical, limiting the improvement in responsivity. Thus, the responsivity increases due to undercutting the devices does not necessarily make up for the dramatic decrease in actuation bandwidth, as illustrated in Fig. 4(b). Eventually, increasing the undercut will cause the actuation platform to become too long and thin to effectively actuate the silicon nitride ring.

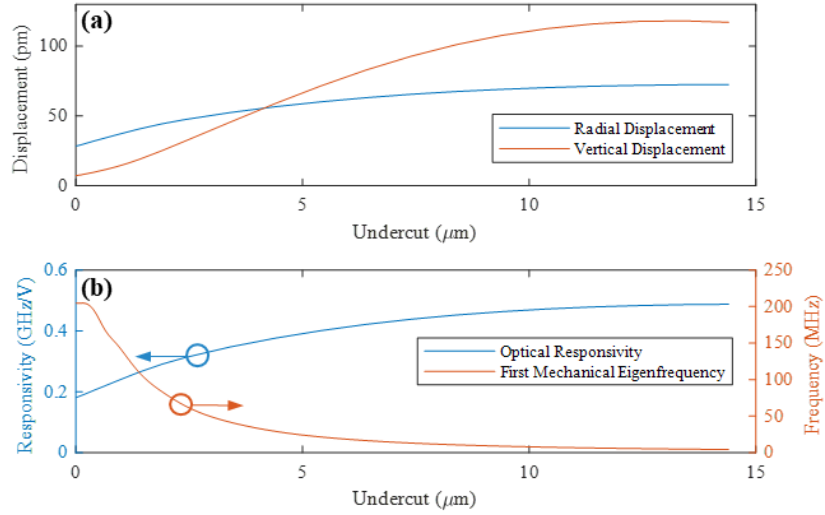

Fig. 4. (a) An axisymmetric finite element analysis of the radial and vertical displacement of the silicon nitride ring as a function of the actuation platform undercut. (b) A plot of the responsivity and the first mechanical eigenmode frequency as a function of the ring modulator undercut.

When the ring is deformed, there is a corresponding stress generated in the ring, as shown in Figs. 3(c) and 3(d), which modifies the refractive index through the photoelastic (or stress-optic) effect by an amount $\Delta n_{ij} = -B_{ijkl}\sigma_{kl}$, where $B_{ijkl}$ and $\sigma_{kl}$ are the stress optic tensor and stress tensor coefficients, respectively. The stress-optic coefficient for silicon nitride is not described in the literature but has been estimated as being close to that of silicon dioxide [8]. Using this estimate, the simulated stress-optic responsivity acts to reduce the total responsivity by approximately half compared to a model with just geometrically induced optomechanical coupling. This does correspond well to the measured responsivity presented later and thus lends some credibility to the use of photoelastic constants similar to silicon dioxide. Of course, these devices could be used to measure the photoelastic coupling constants of silicon nitride by measuring the displacement profile and using that to calibrate the contribution from moving material boundaries, thus allowing the separation of the photoelastic contribution. This will be performed in future work.

Future improvements, such as minimizing the diameter of the central pillar, will give the platform more flexibility and better performance. The central pillar diameter limits the minimum disk actuator and ring diameter, which constrains the platform's ability to achieve high-frequency operation. It also constrains the ability to have a larger undercut of the actuation platform to achieve higher responsivity. As illustrated in Fig. 1(a), the central pillar size is set by the minimum size of vias necessary to make the electrical connections to the "M2" and "M3" aluminum layers. A large, 2 µm central via diameter, which can be seen in Fig. 2(a), was used to ensure a reliable etch through the AlN, and the central pillar was designed with an 8 µm diameter to achieve consistent yield of the electrical connectivity. Reducing this pillar size will allow for a larger undercut or a smaller overall disk in the future adding additional flexibility.

*2.2 Mach-Zehnder Interferometer Design*

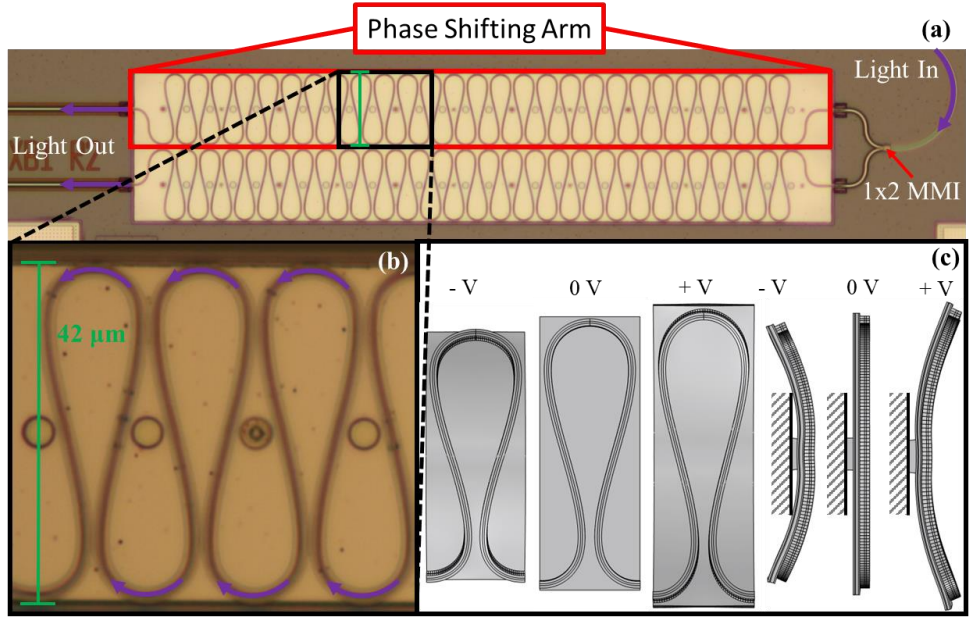

Fig. 5. (a) Microscope image of an MZI modulator. Each arm is 42 µm wide and 480 µm long. The pictured device has 27 periodically repeated bends on each arm. (b) A higher magnification image of a MZI arm. (c) A finite element analysis of the mechanical deformation of one periodic bend of the phase shifting arm. The deformation is scaled from 1 V by 20000x and uses a exaggerated (1 µm thicker) tether for clarity.

The Mach-Zehnder modulator (MZM) demonstrated here is a balanced Mach-Zehnder interferometer with an independently piezo-optomechanically tunable path length in each arm, as shown in Fig. 5(a). This approach allows for broad optical bandwidth operation, in contrast to the ring modulators. Each piezoelectrically tunable phase shifter is composed of a piezoelectric platform that is attached to the substrate along the center across its entire length and fully anchored at the two ends of the platform, such that piezoelectric actuation primarily changes the width of the platform, as shown in Fig. 5(c). The waveguide on each phase shifter meanders back and forth across the width of the piezoelectric platform such that the total optical path length traversed becomes coupled to the width of the platform, allowing a large voltage-dependent phase shift to be accumulated along its path. This is done in such a way as to maximize strain while minimizing optical loss. Figure 5(a), along with a section of a single arm at higher magnification in Fig. 5(b) and a simulation of the piezoelectric deformation of a single period in Fig. 5(c). An optical field at the input goes through a 50:50 MMI splitter, traverses the actuatable sections of the waveguides that comprise the phase shifters, and exits after a 2x2 MMI combiner. The power ratio between the two outputs is determined by the phase difference of the two phase shifting arms. The two phase-shifters are piezoelectrically actuated with equal and opposite voltages, causing one to expand and the other to contract, effectively doubling the responsivity. While for this demonstration, we used two independent voltages that were equal and opposite to achieve the differential actuation of the phase shifters, this could also be achieved with a single drive voltage using the electrical vias to reverse the signal and ground connections to the electrodes for each arm.

The key features of the MZM's cross-sectional geometry are described in Fig. 6(a) AND Table 2 The MZM's were constructed on the same wafers as the ring modulators and thus have the same film thicknesses, which can be seen by comparing to Fig. 2(b) and Table 1.

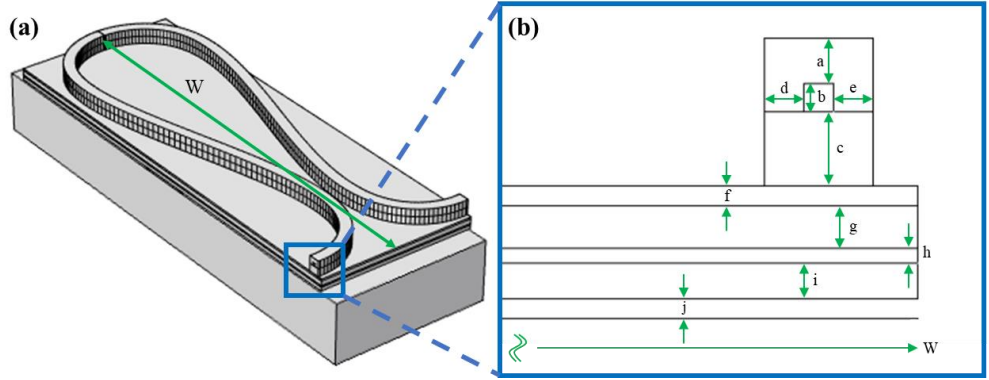

Fig. 6. (a) Model of one periodic bend of the phase shifting arm. (b) Nominal device architecture cross section with labels for critical layer dimensions.

Table 2. Nominal MZM Device Parameters

| Label | a | b | c | d | e | f | g | h | i | j | W |
|---|---|---|---|---|---|---|---|---|---|---|---|
| Material | SiO$_2$ | Si$_3$N$_4$ | SiO$_2$ | SiO$_2$ | SiO$_2$ | Al | AlN | Al | SiO$_2$ | Si | N/A |
| Length (µm) | 0.460 | 0.285 | 0.750 | 0.400 | 0.400 | 0.200 | 0.450 | 0.150 | 0.360 | 0.200 | 42 |

## 3. Results

To demonstrate the viability of our platform, exemplary ring modulator and MZM devices were fabricated at Sandia National Labs' low-volume CMOS production facility, MESA, using the same fabrication tools and at the same time as CMOS production-grade application-specific integrated circuits (ASICs) were being manufactured. The devices were characterized electrically and optically. For all the testing described below, near-visible light from a narrow-linewidth, tunable external cavity diode laser (765 nm - 781 nm) was coupled on and off the chip from a single-mode cleaved fiber using on-chip grating couplers fabricated with the devices. Electrical connection to the chip was made through RF probes in a ground-signal-ground configuration, matched with aluminum pads on the chip.

### 3.1 Ring Modulators

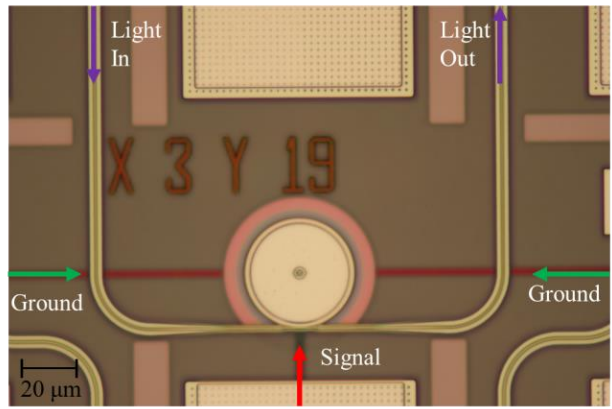

Fig. 7. Microscope picture of a 40.8 µm diameter ring modulator coupled to straight waveguide. GSG pads allow for electrical coupling. Silicon nitride ring with 800 nm width and 285 nm height.

The measured ring modulators were constructed to have a 19.8 µm outer radius, silicon nitride ring (ring width of 0.8 µm, thickness of 0.285 µm) on a 20.4 um diameter AlN actuator platform, as illustrated in Fig. 7. The particular devices discussed here were selected because they exhibited the highest quality factor (lowest scattering losses), though other devices with slightly lower quality factors exhibited on-resonance extinction ratios in excess of 20 dB, due to better phase matching between the ring resonators and coupling waveguides. To investigate differences in piezo-optomechanical actuation, ring resonator modulators were characterized both before and after "release"; i.e., removal of the sacrificial amorphous silicon release layer, as shown in Fig. 1(a). The devices are released by exposing the buried amorphous silicon layer to a xenon difluoride etchant after it is exposed by the plasma etch that defines the disk actuator body, as illustrated in Fig. 1(a). Since the $XeF_2$ etch exhibits reasonable selectivity for a-Si relative to the SiO cladding, the cladding was not substantially removed or roughened and the ring resonators exhibited high optical quality factors even after release, with the device described here exhibiting no reduction in quality factor. Mechanically, and thus piezo-optomechanically, the devices behave very differently before and after release, which will be discussed in detail, below.

We first discuss the non-undercut devices. Fig. 8(a) shows the broad transmission spectrum measured through an integrated bus waveguide, as shown in Fig. 7. Fig. 8(b) shows the transmission spectrum of a single, deeply coupled (-13 dB) resonance with a loaded quality factor of 688,000, (the resonance is overcoupled, yielding an intrinsic Q of ~1.5 million) which yields a loaded half-linewidth of 279 MHz. The responsivity, df/dV, was determined by locking the tunable laser to the optical resonance while applying a sawtooth waveform with 100 ms period to the piezoelectric actuator; a wavemeter with 4 MHz resolution and 3 ms measurement time was used to measure the resulting laser frequency at many points throughout the period. The devices have a highly linear and hysteresis-free voltage-to-frequency responsivity of 200 MHz/V, as shown in Fig. 8(e). The upper horizontal axis of Fig. 8(b) gives the voltage required (relative to zero detuning) to change the transmission via piezo-optomechanical tuning using the measured responsivity. With an initial insertion loss of -13dB, the modulator can be tuned to an extinction ratio of -3dB with an applied voltage of 2 V.

This responsivity described above corresponds to a relative optical path length shift of $5.2 \times 10^{-7}$/V, which is larger than previously demonstrated in electro-optically tuned AlN ring resonators, $3.44 \times 10^{-7}$/V [8] and $3.9 \times 10^{-7}$/V [9]. When compared to electro-optic tuning of AlN, our modulator's higher quality factors and larger relative optical path length shift per volt allows for significantly larger signal modulation per volt in the near-visible than has been previously demonstrated by related techniques.

The non-undercut ring modulators have relatively large effective bandwidths. The small signal response was measured with a vector RF network analyzer, with one port applying an input RF electrical signal while the optical response was measured using a fast photodetector. As shown in Fig. 8(c), the first mechanical eigenfrequency occurs at 204 MHz, which is expected to allow switching in a time of approximately 5 ns. Figure 8(d) shows the result of applying a fast rising-edge voltage waveform via an arbitrary waveform generator; it shows that the device can be switched between two states with a contrast ratio of 13 dB in approximately 4 ns with an 8 V swing. This is presumably at or near the limit set by the resonant response from the first mechanical eigenmode, but it is also the lower limit of the arbitrary waveform generator and thus was not tested at faster switching times.

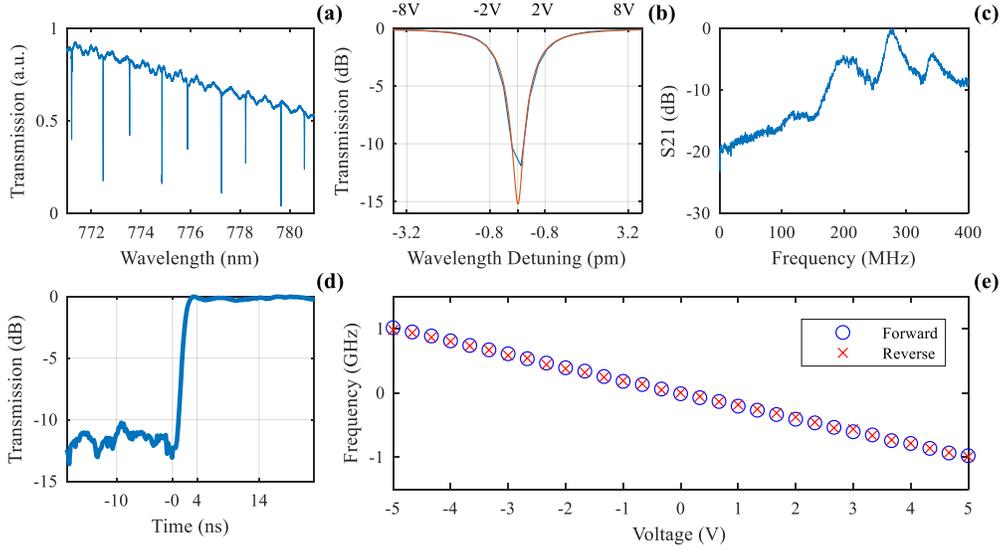

Fig. 8. (a) Transmission spectrum of a non-undercut ring modulator. The actuation platform is 40.8 µm in diameter and has no undercut. Two mode families of different radial orders are clearly present. (b) The measured (blue) transmission spectrum is fitted (orange) using coupled mode theory to determine the intrinsic quality factor and waveguide coupling rate. The lower x-axis is the wavelength detuning of the laser; the upper x-axis is the corresponding calculated voltage required to change the transmission to the value associated with that detuning. (c) The $S_{21}$ response measured using a vector network analyzer and high frequency photodetector. (d) The optical response of a non-undercut ring resonator to a 4ns rise time, 8 $V_{pp}$ electrical pulse average over 16 pulses. (e) Responsivity of the ring modulator in response to a 10 $V_{pp}$ sawtooth waveform. 0.2 GHz/V responsivity was measured.

Figure 9(a) shows the transmission spectrum for a similar ring modulator after the device's sacrificial release layer has been etched away. Releasing the ring modulator significantly improves responsivity while preserving the optical quality factor, despite slight roughening of the $SiO_2$ cladding during the etch, which we observed with a scanning electron microscope; the release also significantly improves responsivity. Figure 9(b) shows the undercut ring device has a loaded quality factor of 792,000. Assuming the resonance is over-coupled to the waveguide yields an intrinsic quality factor 1.8 x $10^6$; there is some ambiguity as to the loading of this resonance, since it is nearly critically coupled. However, a less deeply loaded doublet, one azimuthal mode order higher, exhibits an intrinsic Q of 2.2 x $10^6$ after fitting to the appropriate coupled-mode theory model and accounting for loading and coherent back-scattering [42]. Figure 9(e) shows the responsivity of the modulator, 0.48 GHz/V, which is a 2.4x increase relative to the non-undercut device and corresponds to an optical path length shift of 1.25 x $10^{-6}$/V. This comes at the cost of significantly reducing the frequency at which the first mechanical eigenfrequency occurs, 7.26 MHz as seen in Fig 8(c), compared to 204 MHz for the non-undercut device. Figure 9(d) shows the undercut device switching from maximum extinction to 90% of the modulator's maximum transmission in approximately 100ns. The path length shift of both undercut and non-undercut devices are significantly higher than previously demonstrated modulators using AlN piezoelectric tuning of silicon nitride, which showed a relative shift of 2.15 x $10^{-7}$/V when undercut to a mechanical eigenfrequency of 1.1 MHz [22] and a relative shift of 1.25 x 10^-7/V when actuating a completely buried ring resonator [23, 24].

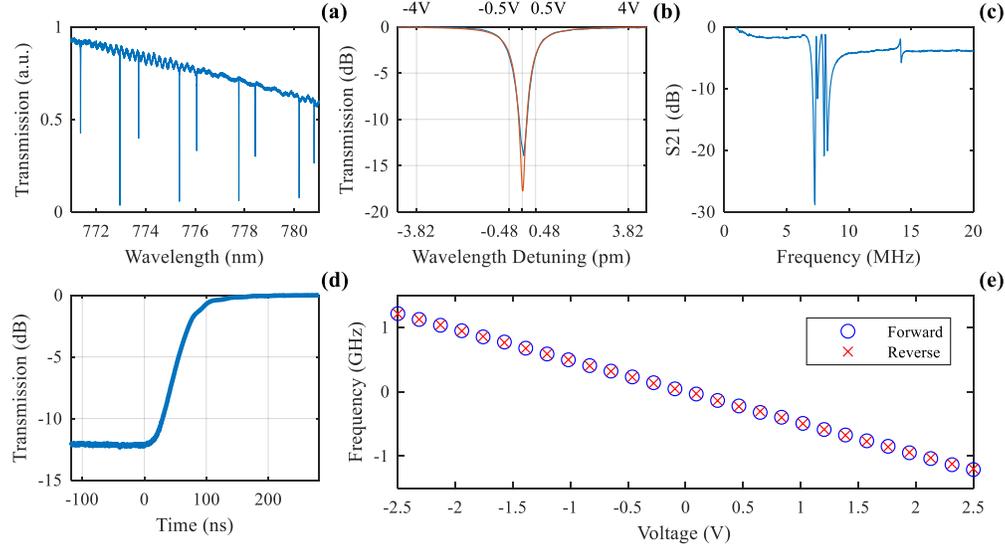

Fig. 9. (a) Transmission spectrum of a released ring modulator. The actuation platform is 40.8 µm in diameter and is undercut by 12.4 µm. Two mode families of different radial orders are again present. (b) Measured (blue) and fit (orange) lineshape for a single, tested resonance. The lower x-axis is the wavelength detuning of the laser; the upper x-axis is the corresponding calculated voltage required to change the transmission to the value associated with that detuning. (c) The $S_{21}$ response for the undercut ring modulator measured using a vector network analyzer and photodetector. The first mechanical eigenfrequency appears at 7.26 MHz. (d) The optical response of an undercut ring resonator to a 200 ns rise time, 4 $V_{pp}$ electrical pulse average over 16 pulses. (e) Responsivity of the ring modulator in response to a 5 $V_{pp}$ slow saw-wave. 0.48 GHz/V responsivity was measured.

Though near-visible electro-optic modulators using AlN as the waveguiding material have been demonstrated with higher bandwidths (approximately 10 GHz) than our piezoelectric modulation [8], realistic constraints on voltage and modulation depth make it impossible to achieve those bandwidths at near-visible wavelengths and shorter. The bandwidth of a resonant modulator, $f_{3dB}$, is related to the photon lifetime

$$\tau_{ph} = \frac{\lambda Q}{2\pi c}; \quad \frac{1}{f_{3dB}} = 2\pi \tau_{ph}$$

Previous demonstrations of electro-optic modulators in AlN achieved very low-Q resonators (~10,000) with an estimated responsivity of 0.28GHz/V [8]. The low quality factors did allow for fast modulation, but the modulation depth demonstrated was not sufficient for many information processing applications and invokes significant transmitter power penalties in others. This is because, given the low Q and responsivity, they would require approximately 400 V to achieve the same modulation depth we have demonstrated here. To achieve greater modulation depths with practical RF voltages, the Q would have to be significantly increased, as the responsivity is mostly fixed by the AlN electro-optic coefficient and minimum necessary cladding thicknesses. This would, in turn, decrease the achievable bandwidth. In contrast, our non-undercut modulators have a bandwidth-limiting mechanical eigenfrequency of 204 MHz as currently constructed and shown in Fig. 8(c). A ring resonator with an intrinsic quality factor of 1.5 x 10^6 would have a photon lifetime limited 3 dB frequency cutoff of 0.5 GHz when critically coupled to a waveguide. Reducing the diameter of our piezoelectric modulators by a factor of two would raise the mechanical eigenfrequency to approximately this frequency, allowing us to reach the photon lifetime limited modulation bandwidth with a large modulation depth and low switching voltage at the same time. We could also continue to decrease the resonator diameter, which would continue raising the mechanical bandwidth limit, while also raising the photon lifetime bandwidth limit because we would quickly reach the radiation limit of Q at small diameters. Thus, we expect that this platform could achieve significantly larger bandwidths and still maintain practical switching voltages.

One of the advantages of using aluminum nitride for piezoelectric switching and modulation is its extremely high resistivity, which results in very low leakage currents and power dissipation. For the ring resonator modulators described here, we measure a room temperature leakage current of 0.01 nA at 5 V, which corresponds to a device resistance of 500 GΩ and a resistivity of $1.5 \times 10^{11}$ Ω-cm. To maintain a non-undercut device in a state 10 dB from its maximum extinction requires a steady-state power dissipation of 8 pW at 2 V. For comparison, the non-undercut ring modulator would require 0.62 mW to maintain the same switched state using thermo-optic tuning and a typical pathlength tuning coefficient of $1.7 \times 10^{-6}$ /mW [14] for clad $Si_3N_4$ ring resonators [3].

The energy per bit required to change the state of the modulator is due to the work done by the power supply to charge the capacitance of the device. The capacitance of the device is $C = \varepsilon A/d$, where $\varepsilon$ is the permittivity of aluminum nitride (the dielectric constant of AlN is approximately 10), A is the area of the actuator, and d is the separation of the metal layers, which in this case is the AlN thickness of 450 nm. This yields a capacitance of 0.26 pF. The energy to charge this capacitance is given by $U = 1/2 * CV^2$, which for the demonstrated non-undercut modulator switching between maximum extinction and 10 dB higher power with a 2 V driving voltage yields an energy of 520 fJ/bit.

*3.2 Mach-Zehnder Modulators*

The exemplary MZM device consists of 27 periodically repeated bends across the 42 µm width and 480 µm length of each phase shifting arm, as shown in Figs. 5(a) and 5(b). Measurements with an RF network analyzer, the results of which are shown in Fig. 10(a), shows that the system could potentially be operated with a rise time near 250 ns, but we were limited to demonstrating switching with a 750 ns rise time, as shown in Fig. 10(b), due to the bandwidth limitations of the high-voltage amplifier used to boost the voltage from an arbitrary waveform generator. With a drive voltage amplitude of 20 V (40 V peak-to-peak) and a total waveguide length of 3 mm per arm, the tested device achieved a differential π phase shift between the two arms. This corresponds to an $LV_\pi$ of 24 V·cm and a relative optical path length shift of $1.62 \times 10^{-6}$. The loss per waveguide wrapping period was found to be 0.07 dB via measurement of devices of increasing numbers of bending periods, which corresponds to approximately 1.89 dB loss in each arm of the 27-period measured device. While this is not a particularly low loss per period, limited optimization was conducted to minimize the bending loss, since the main focus of this work was to assess the achievable phase shift per volt. With further optimization, it should be possible to approach the propagation loss demonstrated in the ring resonators, around 0.4 dB/cm, which would reduce the loss to around 0.12 dB loss over the length of the MZI arm. Recent work in electro-optic modulation of AlN waveguides have demonstrated an $LV_\pi$ of 72.2 V·cm at a wavelength of 1064 nm, but still experience rapidly increasing loss at shorter wavelengths with a propagation loss of 3.7 db/cm at 782 nm [10]. In the case of a modulator operating near 780 nm, the $LV_\pi$ of 72.2 V·cm in [10] would scale to 53 V·cm; if limited to 20 V drive amplitude as in this work (40 V applied), the modulator would need to be 1.3 cm and incur a loss of 4.8 dB, given the propagation loss of 3.7 dB/cm. On the other hand, if the modulator demonstrated in this work had its bending loss reduced below the propagation loss limit, the same device length of 1.3 cm would incur a loss of only 0.52 dB and would have a $V_\pi$ of 18 V.

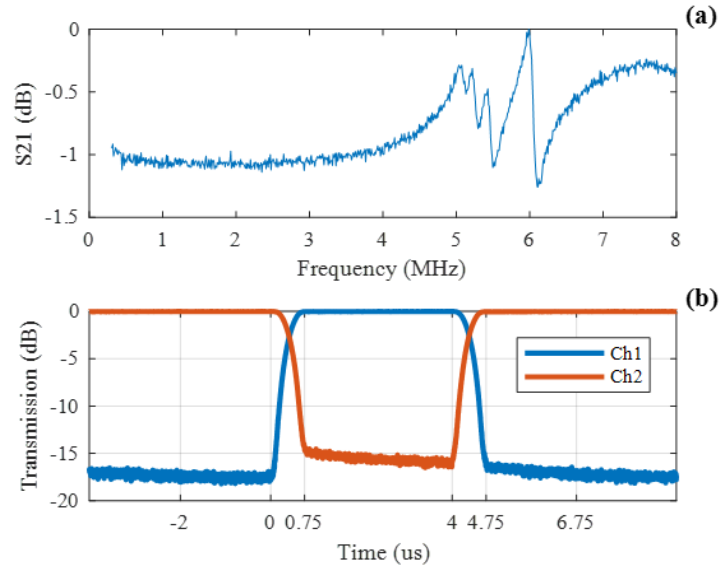

Fig. 10. (a) The $S_{21}$ response for a single arm measured with a vector network analyzer and photodetector. The first mechanical eigenfrequency appears at 5.05MHz. (b) A 40 $V_{pp}$ pulse with a 750 ns rise time (amplifier limited) is applied to two arms of a MZI. Opposite voltages applied to each channel. The actuation platform for each waveguide arm is 42um wide, 18.5um undercut, and 480um long. Data is averaged over 16 pulses.

The switching results of the MZI devices are comparable to electrostatically actuated optical switching circuits [2, 4, 5, 43] when evaluating switching speed and voltages. Designs incorporating more complicated MMI structures have demonstrated better results [1, 4], but such designs are often difficult to fabricate.

The energy and power necessary to operate the MZM are the same as those discussed for the ring resonator. Average leakage current from a fabricated MZM arm at 20 V was 0.5 nA at room temperature, which corresponds to a device resistance of 40 GΩ and a sheet resistance of 190 GΩ-cm. A π phase shift for quasi-DC operation of these piezo-optomechanical MZMs requires only 20 nW to maintain at room temperature, which is 10 times lower than demonstrated PZT strain-based phase modulators at room temperature while achieving similar bandwidth and responsivity [26, 27]. Using the previous relative shift per power for thermo-optic tuning, $1.7 \times 10^{-6}$/mW [14], the switched state would require 38 mW to maintain. The capacitance of each MZM arm is 4.1 pF. The energy to charge this capacitance at a 40 V switching signal yields an energy of 6.6 nJ/bit.

*3.2 Alternative Phase Shifter Design*

The piezo-optomechanical phase shifters and associated Mach-Zehnder modulators described above were designed based on being the non-resonant analogues of the ring resonators, having large piezoelectric actuator platforms and thus large responsivity that allows for low-voltage operation in a compact space. An alternative approach is to use a straight waveguide (no bends) that is strained and has its cross-sectional dimensions modified by the piezoelectric modulation. Figures 11(a) and 11(b) shows an example and finite element model results of this type of piezo-optomechanical waveguide phase shifter, where the stress-optic tensor coefficients of silicon dioxide have again been used for those of silicon nitride. As shown in Fig. 11(c), the responsivity is a strong function of the width, favoring waveguides wider than the single-mode cut-off. This is somewhat advantageous, since the use of a waveguide much wider than the single-mode cutoff allows for reduced scattering losses from sidewalls, which in turn allows for centimeter-scale propagation lengths that lead to low values of $V_\pi$. The responsivity is significantly smaller than the previously presented MZM design, requiring larger lengths for practical voltage limits; however, in the regime where the responsivity is largest (widest waveguides), the stress-optic effect is dominant, which leaves open the possibility that silicon nitride could actually have a significantly larger responsivity. Furthermore, this straight waveguide design does not suffer from bending losses and has a comparatively high mechanical eigenfrequency, leading to a large bandwidth. For example, the first mechanical eigenfrequency appears at 233MHz for 5 µm wide waveguide actuated by a 10 µm wide actuation platform. While there is some variation in the eigenfrequency as a function of width, they maintain high fundamental eigenfrequencies across the range described in Figure 11. Devices of this design offer the benefits of the ring modulators demonstrated, such as low voltage operation and high bandwidth, but can be done with broadband optical operation, albeit at the expense of relative compactness. These devices will be explored in more detail in future work.

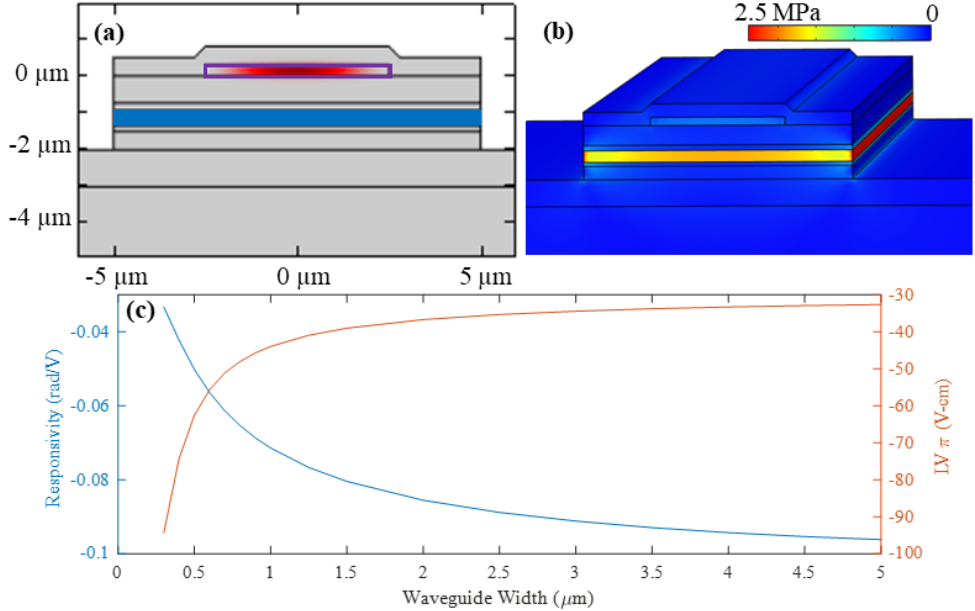

Fig. 11. (a) A finite element analysis of the TE00 optical mode, with optical power shown in red, of a 5 µm wide silicon nitride waveguide (purple outline) being piezo-optomechanically actuated by a 10 µm wide AlN platform (blue). (b) Von Mises stress of the same modulator at 1 V shown in 3D. (c) A graph of the optical responsivity and $LV_\pi$ as a function of the waveguide width.

### 3.3 Cryogenic Characterization

To test the platform's compatibility with cryogenic environments, the ring modulator devices were tested optically and electrically in a closed loop cryostat. Figure 12 shows the performance characteristics of a ring resonator device tested at 7 K. The piezoelectric responsivity of 0.26 GHz/V did not significantly deviate from room temperature and is free of hysteresis. This is quite unlike semiconductor-based modulators based on carrier dispersion that require significant device modification to operate at cryogenic temperatures due to large variations in carrier concentration [44]. A large advantage of this system is that it is extremely stable to temperature fluctuations at cryogenic temperatures, due to the dramatic decreases in thermal expansion and thermo-optic coefficients at these temperatures. Previous work has demonstrated the decreasing thermo-optic coefficient of PECVD silicon nitride resonators at temperatures approaching absolute zero [14]. At 18 K, the change in refractive index with temperature is below 300 ppb/K for silicon dioxide and silicon nitride. In addition, resistance measurements taken at cryogenic temperatures of a ring modulator showed a resistance greater than 20 TΩ (6.1 TΩ-cm), decreasing the power dissipation required to operate the device by a factor of 40 compared to room temperature, which is crucial in a cryogenic environment where all dissipated heat must be removed by the cooling system to maintain the operating temperature. In particular, this would result in dissipating only 200 femtowatts to hold the non-undercut resonator at 10 dB contrast from its maximum insertion loss state (2 V applied).

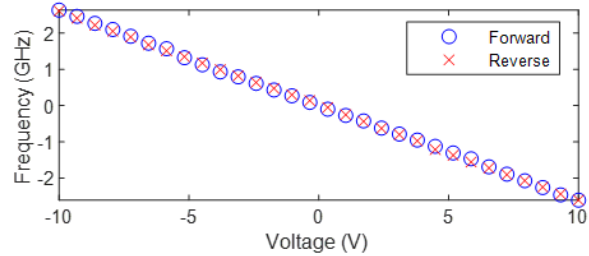

Fig. 12. Wavemeter measured responsivity of a ring resonator in response to a 20 $V_{pp}$ slow saw-wave at 7 K.

The fact that the piezoelectrically actuated devices exhibit high performance at cryogenic temperatures makes them suitable candidates for cryogenic photonic integrated circuit components, especially in the near-visible and visible spectrum. This should be contrasted with conventional thermo-optic tuning at these temperatures. Though the specific heat of silicon nitride decreases with temperature, it does not fully compensate for the greatly reduced thermo-optic coefficient in regards to the necessary on-chip power dissipation to achieve a given phase shift [14]. These temperature-dependent material parameters decrease the efficacy of thermo-optic tuning, greatly increasing the heat generated during tuning and thus the cooling power required of the cryostat. For example, a thermo-optic phase modulator made from a 1 cm long silicon nitride waveguide starting at 7 K would need to increase in temperature to approximately 41 K to create a π phase shift. If we use the approximate relative path length shift of 1.14 x $10^{-6}$/mW for thermo-optic tuning at 18 K [14], then a π phase shift in a 10 cm length would require 19.7 mW. In addition, a thermally tuned cryogenic photonic integrated circuit would need to accommodate significant local temperature variations, making high density of channels in these photonic integrated circuits extremely impractical.

## 4. Conclusion

In this work, we have presented a viable platform for phase and amplitude modulation of silicon nitride photonic integrated circuits. The platform offers a solution with low optical loss and low power consumption for applications with operating wavelengths over the transparency window of silicon nitride, particularly the near-visible and visible portions of the electromagnetic spectrum, and the platform works without any performance degradation from room temperature to 7 K, with the expectation that the performance would remain unchanged down to arbitrarily low temperatures. This platform was fabricated at a production CMOS foundry providing a clear path for truly scalable photonic integrated circuits for visible wavelength and cryogenic applications, including integrated CMOS electronics. It should be possible to substitute the silicon nitride waveguide layer for an alumina ($Al_2O_3$) layer to extend the performance into the blue and ultraviolet wavelength regimes [45].

One advantage of our system compared to similar systems is the decoupled nature of the piezoelectric actuation and optical properties of the system, allowing for independent optimization of each. In the future, we hope to improve the piezoelectric tuning of the system by alloying aluminum nitride with scandium nitride. AlN-ScN alloy thin films have demonstrated up to 5x improvement of the piezoelectric coefficients [46]. This would allow for lower drive voltages or the actuation of stiffer designs for faster modulation speeds. Also, further reduction in the optical losses should be possible with continued fabrication optimization. Even without further optimization, the exceptionally low on-chip power dissipation, cryogenic compatibility, stable operation, and large optical transparency window will be transformative for many integrated photonic circuit applications.

## 5. Acknowledgments

This work was supported by the Laboratory Directed Research and Development program at Sandia National Laboratories, a multimission laboratory managed and operated by National Technology and Engineering Solutions of Sandia, LLC., a wholly owned subsidiary of Honeywell International, Inc., for the U.S. Department of Energy's National Nuclear Security Administration under Contract No. DE-NA-003525. This paper describes objective technical results and analysis. Any subjective views or opinions that might be expressed in the paper do not necessarily represent the views of the U.S. Department of Energy or the United States Government.